\documentclass[12pt]{amsart}
\usepackage{amsmath, amssymb,latexsym }
\usepackage{verbatim}

\renewcommand{\phi}{\varphi}
\renewcommand{\epsilon}{\varepsilon}
\renewcommand{\emptyset}{\varnothing}

\numberwithin{equation}{section}

\newcommand{\e}{\varepsilon}
\newcommand{\HH}{\mathcal H}
\newcommand{\Ss}{\mathbb S^2}
\renewcommand{\le}{\leqslant}
\renewcommand{\ge}{\geqslant}
\newcommand{\ci}[1]{_{{}_{\scriptstyle{#1}} } }
\newcommand{\E}{\mathbb E}
\newcommand{\PP}{\mathbb P}

\newtheorem{thm}[equation]{Theorem}
\newtheorem{lemma}[equation]{Lemma}

\newtheorem{cor}[equation]{Corollary}

\newtheorem{claim}[equation]{Claim}
\newtheorem{rem}[equation]{Remark}

\title[On the number of nodal domains]{On the number of nodal domains \\
of random spherical harmonics}

\author{Fedor Nazarov}
\address{Department of Mathematics\\
Michigan State University\\
East Lansing, MI 48824\\
USA}
\email{fedja@math.msu.edu}

\author{Mikhail Sodin}
\address{School of Mathematical Sciences\\
Tel Aviv University\\
Tel Aviv 69978\\
Israel}
\email{sodin@post.tau.ac.il}

\begin{document}

\begin{abstract}

Let $N(f)$ be a number of nodal domains of a random Gaussian
spherical harmonic $f$ of degree $n$. We prove that as $n$ grows
to infinity, the mean of $N(f)/n^2$ tends to a positive constant
$a$, and that $N(f)/n^2$ exponentially concentrates around $a$.

This result is consistent with predictions made by Bogomolny and Schmit
using a percolation-like model for nodal domains of random Gaussian
plane waves.
\end{abstract}

\maketitle

\section{Introduction}\label{section_intro}

Let $\HH_n$ be the $2n+1$-dimensional real Hilbert space of
spherical harmonics of degree $n$ on the $2$-dimensional unit
sphere $\Ss$ equipped with the $L^2(\Ss)$ norm. For $f\in\HH_n$,
put $Z(f)=\{x\in\Ss\colon f(x)=0\}$. Let $N(f)$ be the number of
connected components of $Z(f)$. The famous Courant nodal domain
theorem \cite[Chapter~VI, \S~6]{CH} states that $N(f)\le (n+1)^2$
for all $f\in\HH_n$. On the other hand, H.~Lewy~\cite{Lewy} showed
that no non-trivial lower bound is possible: one can find
spherical harmonics $f$ of arbitrarily large degree with $N(f)\le
3$. The question we want to discuss here is: \textit{What is the
``typical'' value of $N(f)$ when the degree $n$ is large}? To give
the word ``typical'' a precise meaning, let us consider the random
spherical harmonic
$$
f=\sum_{k=-n}^{n}\xi_k Y_k
$$
where $\xi_k$ are independent identically distributed Gaussian
random variables with $\E \xi_k^2 = \tfrac1{2n+1}$ and $\{Y_k\}$
is an orthonormal basis of $\HH_n$, so $ \E\|f\|\ci{L^2(\Ss)}=1$.
It is not hard to see that $f$ (as a random function) does not
depend on the choice of the basis $\{Y_k\}$ in $\HH_n$.

The same question can be raised in other instances of smooth
random functions of several real variables, e.g., for random
trigonometric polynomials of large degree $n$. We are not aware of
any {\em rigorous} treatment of this question, though we know two
encouraging attempts to tackle it in very different contexts. In
the paper \cite{Sw} (motivated by some engineering problems),
Swerling estimated from below and from above the mean number of
connected components of the {\em level lines} $Z(t, f) = \{ f=t\}$
of a random Gaussian trigonometric polynomial $f$ of two variables
of given degree $n$. His method is based on estimates of the
integral curvature of the level line $Z(t, f)$. The estimates are
rather good when the level $t$ is separated from zero, but as
$t\to 0$ they are getting worse and, unfortunately, give nothing
when $t=0$.

A few years ago Blum, Gnutzmann, and Smilansky \cite{BGS} raised
a question about the distribution of the number of nodal domains
of high-energy eigenfunctions. In the ergodic case, in accordance
with Berry's ``random wave conjecture'', they suggested to find
this distribution for Gaussian random plane waves and performed
the corresponding numerics. To compute this distribution,
Bogomolny and Schmit suggested in \cite{BS} an elegant
percolation-like lattice model for description of nodal domains of
random Gaussian plane waves. It agrees well with numerics, but
completely ignores the correlation between values of the random
function $f$ at different points, and apparently it will be very
difficult to make it rigorous.

In this note, we will show that, in accordance with one of
the Bogomolny and Schmit predictions, $\E N(f)/n^2$ tends to a positive
limit $a$ when $n\to\infty$. Moreover, we show that the random variable
$N(f)/n^2$ exponentially concentrates around $a$:
\begin{thm}\label{thm.main}
There exists a constant $a>0$ such that, for every $\e>0$, we have
$$
\PP\left\{\left|\frac{N(f)}{n^2}-a\right|>\e\right\}\le
C(\e)e^{-c(\e)n}
$$
where $c(\e)$ and $C(\e)$ are some positive constants depending on
$\e$ only.
\end{thm}

\begin{rem}[Sharpness of Theorem~\ref{thm.main}]
{\rm The exponential decay in $n$ in Theorem~\ref{thm.main} cannot
be improved: in Section~\ref{section_example} we show that,} given
a positive and arbitrarily small $\kappa$, $\displaystyle
\PP\left\{ N(f) < \kappa n^2 \right\} \ge e^{-C(\kappa)n}$. {\rm
On the other hand, our proof of Theorem~\ref{thm.main} gives a
very small value $c(\epsilon) \gtrsim \epsilon^{15}$ and it would
be nice to reduce the power $15$ of $\epsilon$ to something more
reasonable.}
\end{rem}

\begin{rem}{\rm The model proposed by Bogomolny and Schmit
also predicts that the variance of the random variable $N(f)$
grows with $n$ as $bn^2$ with some constant $b>0$. }
\end{rem}

\begin{rem}
{\rm For any spherical harmonic $f\in\HH$, the total length of its
nodal set $Z(f)$ does not exceed ${\rm Const}\, n$. Therefore,
Theorem~\ref{thm.main} yields that, for a typical spherical
harmonic, most of its nodal domains have diameters comparable to
$1/n$. }
\end{rem}

The proof of Theorem~\ref{thm.main} goes as follows:

\smallskip\par\noindent{\bf I}. First, we prove the lower bound $\E N(f) \ge
{\rm const}\, n^2$. This part of the proof is rather
straightforward and short.

\smallskip\par\noindent{\bf II}. Then we prove the exponential concentration of the
random variable $N(f)/n^2$ around its median. This part is based
on two ingredients:

\par\noindent{\bf (i)} the uniform lower continuity of the
functional $f\mapsto N(f)$ with respect to the $L^2$-norm outside
of an exceptional set $E\subset \HH$ of exponentially small
measure;

\par\noindent{\bf (ii)} Levy's concentration of measure principle.

\smallskip\par\noindent{\bf III}. In the third part, we prove
existence of the limit $\displaystyle \lim_{n\to\infty} \E
N(f)/n^2$. In this part, we use existence of the scaling limit for
the covariance function $\E \big\{ f(x)f(y) \big\}$.

\smallskip
Note that in the proof of Theorem~\ref{thm.main} we use only
relatively simple tools from the classical analysis, which we
believe may work in a more general setting of random functions of
several real variables, while it seems that the Bogomolny-Schmit
model is essentially a two-dimensional one.

\subsection*{Notation} Throughout the paper, we denote by $c$
and $C$ positive numerical constants whose values may vary from
line to line. As usual, the constants denoted by $C$ are large,
while the ones denoted by  $c$ are small. In the cases when we
need to fix the value of some constant, we assign to it a certain
index, for instance, $c_0$ and $C_0$.

By $\mathcal D(x, r)$ we denote the spherical disk of radius $r$
centered at $x$, by $D(y, t)$ we denote the Euclidean disk of
radius $t$ centered at $y$.

By $\sigma$ we denote the spherical area measure with
normalization $\sigma (\Ss) = 1$, and by $m$ we denote the
(Euclidean) area measure on the plane.

By $\|\, \cdot\,\|$ we always mean the $L^2(\Ss)$-norm.

Given a set $K$, we denote by $K_{+d}$ the $d$-neighbourhood of
$K$. We apply this notation both to subsets of $\HH_n$ and the
$L^2$-distance, and to subsets of $\Ss$ and the usual spherical
distance.

Notation $A \lesssim B$ and $ A \gtrsim B $ means that there exist
positive numerical constants $C$ and $c$ such that $A\le C\cdot B$
and $ A \ge c \cdot B $. If $A\lesssim B$ and $A\gtrsim B$
simultaneously, then we write $A\simeq B$. Notation $A \ll B$
stands for ``much less'' and means that $A \le c \cdot B$ with a
very small positive $c$; similarly, $A \gg B$ stands for ``much
larger'' and means that $A\ge C\cdot B$ with a very large positive
$C$.

\subsection*{Acknowledgments} We learned  about the
problem considered in this note and about the works~\cite{BGS, BS}
from Ze\'ev Rudnick. We thank him as well as Leonid Polterovich,
Boris Tsirelson and Steve Zelditch for very helpful discussions.

\section{Main tools}\label{section_tools}

\subsection{Spherical harmonics}

We shall need a few standard facts about spherical harmonics of
degree $n$. Most of them can be derived either from the fact that
they are eigenfunctions of the Laplacian on the sphere
corresponding to the eigenvalue $n(n+1)$ or from the fact that
they are traces of homogeneous harmonic polynomials of degree $n$
on the unit sphere. Everywhere below we assume that $n\ge 1$.

\begin{claim}\label{claim0}
The scalar product in the Hilbert space $\HH_n$ is invariant under
rotations (and reflections) of the unit sphere. As a consequence,
the distribution of the random spherical harmonic $f$ is also
rotation invariant.
\end{claim}


\begin{claim}\label{claim2}
For any $f\in\HH_n$ and any point $x\in \Ss$, we have
$$
\begin{aligned}
|f(x)|^2 & \lesssim n^2\int_{\mathcal D(x,1/n)}f^2\,;
\\
|\nabla f(x)|^2 & \lesssim  n^4\int_{\mathcal D(x,1/n)}f^2\,;
\\
|\nabla\nabla f(x)|^2 &\lesssim  n^6\int_{\mathcal D(x,1/n)}f^2\,.
\end{aligned}
$$
\end{claim}

\begin{claim}[Length estimate]\label{claim3}
For any $f\in \HH_n$ that is not identically $0$, the total length
of $Z(f)$ does not exceed $Cn$.
\end{claim}

The next claim follows from the classical Faber-Krahn inequality:

\begin{claim}[Area estimate]\label{claim4}
For any connected component $\Omega$ of $\Ss\setminus Z(f)$, we
have $\operatorname{Area}(\Omega) \gtrsim   n^{-2}$.
\end{claim}

Next, we bring several classical facts about the Legendre
polynomials $\displaystyle P_n(x) = \frac1{2^n n!} \frac{d^n
(x^2-1)^n}{dx^n}$. Note that $P_n(1)=1$.

\begin{claim}\label{claim5}
The function $\displaystyle Y_0(\phi, \theta) = \sqrt{2n+1}
P_n(\cos\theta) $ is a spherical harmonic of degree $n$ with $\|
Y_0 \| =1 $. Here $(\phi, \theta) $ are the longitude and the
co-latitude on the sphere $\Ss$.
\end{claim}
The function $Y_0$ is called a {\em zonal spherical harmonic}.

\begin{claim}\label{claim_zonal} There exists a positive constant
$c_0$ such that
\[
Y_0^2 + \frac1{n^2} |\nabla Y_0|^2 \ge c_0^2
\]
everywhere on $\Ss$.
\end{claim}
This claim is a combination of two classical results:

\smallskip\par\noindent{\bf (i)} The function
$x\mapsto n(n+1) P_n^2 + (1-x^2)(P_n')^2$ increases on $[0, 1]$
(this is Sonine's theorem~\cite[Section~7.3]{Szego}).

\smallskip\par\noindent{\bf (ii)} $|P_{2m}(0)| = g_m$ and
$|P_{2m-1}'(0)|= 2m g_m$, where $g_m = \tfrac{1\cdot 3\cdot\,...\,
\cdot (2m-1)}{2 \cdot 4 \cdot \, ...\, \cdot 2m}$. This follows,
for instance, from the recurrence relations
\[ (n+1)P_{n+1}(x) = (2n+1) x P_n(x)  - nP_{n-1}(x) \] and \[
P_{n+1}'(x) = xP_n'(x) + (n+1) P_n(x) \] \cite[Chapter~VII,
\S~3]{CH}. Recall that by Wallis' formula, $g_m \ge
\tfrac{c}{\sqrt{m}}$.

\begin{claim}[Reproducing kernel in $\HH_n$]\label{claim5.5}
\[
\frac{1}{2n+1} \sum_{k=-n}^n Y_k(x) Y_k(y) =
P_n(\cos\Theta(x,y))\,,
\]
where $\Theta(x, y)$ is the angle between the vectors $x, y
\in\Ss$. In particular,
\[
\sum_{k=-n}^n Y_k^2(x) = 2n+1\,.
\]
\end{claim}

The next two facts can be found in Szeg\"o's book \cite{Szego}
(Theorems 6.21.2 and 8.21.6 correspondingly).

\begin{claim}\label{claim6}
Suppose $x_\nu = \cos \theta_\nu$ are zeroes of $P_n$ enumerated
in decaying order:
\[
+1 > x_1 >\, ...\, > x_n > -1, \qquad 0<\theta_1 < \, ...\, <
\theta_n< \pi\,.
\]
Then
\[
\frac{2\nu-1}{2n+1}\pi < \theta_\nu < \frac{2\nu}{2n+1}\pi, \quad
\nu = 1, 2, \, ...\, n\,.
\]
\end{claim}

\begin{claim}[Hilb's asymptotics]\label{claim7}
\[
P_n(\cos\theta) = \left( \frac{\theta}{\sin\theta} \right)^{1/2}
J_0 \big( (n+\frac12)\,\theta \big) + R\,,
\]
where
\[
R =
\begin{cases}
\theta^{1/2} O(n^{-3/2}), & C/n \le \theta \le \pi/2 \\
\theta^2 O(1), & 0\le \theta \le C/n\,,
\end{cases}
\]
and $J_0$ is the zeroth Bessel function.
\end{claim}
Note, that we shall use Claim~\ref{claim6} only for $\nu = 1$ and
$2$, and Claim~\ref{claim7} for $0\le \theta \le C/n$.

\subsection{Probabilistic claims}

We shall also need a few classical facts about the Gaussian random
vectors in spaces of high dimension.

\begin{claim}[Bernstein's concentration of norm]\label{claim8}
$$
\PP\left\{\|f\| >2  \right\}\le e^{-cn}\,.
$$
\end{claim}

The next result follows from the Gaussian isoperimetric lemma
which is due to Sudakov-Tsirelson~\cite{ST} and Borell~\cite{B}:

\begin{claim}[Levy's concentration of Gaussian measure]\label{claim9}
Let $F\subset\HH_n$ be any measurable set of spherical harmonics.
Suppose that the set $F_{+\rho}$ satisfies $\PP(F_{+\rho})<\frac
34$. Then $\PP(F)\le 2 e^{-c\rho^2 n}$.
\end{claim}

\begin{claim}[Independence of $f$ and $\nabla f$]\label{claim10}
If $x\in \Ss$, then $f(x)$ and $\nabla f(x)$ are independent
Gaussian random variables. Also, due to rotation invariance, we
can say that $\E |f(x)|^2=1$, $\E|\nabla f(x)|^2\lesssim n^2$, and
that the distribution of \, $ \nabla f (x)$ is rotation invariant
on the tangent plane $T_x(\Ss)$.
\end{claim}

\section{Lower bound for $\E N(f)$}\label{section_LB}

Here, we show that $\E N(f) \gtrsim n^2$. The proof has two
ingredients: an estimate of the maximum $\displaystyle
\max_{\mathcal D(x, \rho/n)} |f|$ and existence of the ``barrier
function'' $b_x$.

\begin{claim}[Estimate of the maximum]\label{claim11}
Given $\rho>0$, there exists $C_0$ such that, for any $x\in\Ss$,
$\displaystyle \PP \big\{ \max_{\mathcal D(x, \rho/n)} |f|   \ge
C_0\big\} \le \tfrac13$.
\end{claim}

\par\noindent{\em Proof of Claim~\ref{claim11}:} By the
mean-value inequality in Claim~\ref{claim2}, for any spherical
harmonic $f\in \HH_n$ and any $x\in\Ss$, we have
\[
\max_{\mathcal D(x, \rho/n)} f^2 \lesssim n^2 \int_{\mathcal D(x,
(\rho+1)/n)} f^2\,.
\]
Integrating this inequality with respect to $x$ over the sphere,
changing the integration order, and taking into account that $n^2
\sigma \big(\mathcal D(\,\cdot\,, \tfrac{\rho+1}{n}) \big)
\lesssim 1$, we get
\[
\int_{\Ss} \max_{\mathcal D(x, \rho/n)} f^2 \lesssim \int_{\Ss}
f^2\,.
\]
Hence,
\[
\E \big\{ \max_{\mathcal D(x, \rho/n)} f^2 \big\} =  \E \Big\{
\int_{\Ss} \max_{\mathcal D(x, \rho/n)} f^2 \Big\} \lesssim \E
\|f\|^2 = 1
\]
(in the first equation we used the rotation invariance of the
distribution of $f$). Applying Chebyshev's inequality, we get the
estimate. \hfill $\Box$

\begin{claim}[Existence of the barrier]\label{claim12}
There exist positive numerical constants $\rho$ and $c_1$ such
that, for each sufficiently large $n$ and each $x\in \Ss$, there
is a function $b_x\in\HH_n$ with the following properties:
\[
\| b_x \| = 1\,, \qquad b_x(x) \ge c_1 \sqrt{n}\,, \quad
\text{and} \quad b_x\big|_{\partial \mathcal D(x, \rho/n)} \le
-c_1 \sqrt{n}\,.
\]
\end{claim}

\par\noindent{\em Proof of Claim~\ref{claim12}:} If $x$ is the North
Pole, then by Claims~\ref{claim5}, \ref{claim6} and \ref{claim7}
the zonal spherical harmonic $Y_0$ gives us what we need. For
other $x$'s, we just rotate the sphere. \hfill $\Box$

\bigskip\par\noindent{\em Proof of the lower bound for $N(f)$: }
Fix $x\in \Ss$. We have $f=\xi_0 b_x + f_x$ where $\xi_0$ is a
Gaussian random variable with $\E \xi_0^2 = \tfrac1{2n+1}$, and
$f_x$ is a Gaussian spherical harmonic built over the orthogonal
complement to $b_x$ in $\HH_n$ and normalized by $\E \| f_x \|^2 =
\tfrac{2n}{2n+1}$. We choose a Gaussian random variable
$\widetilde\xi_0 $ independent of $\xi_0$ and of $f_x$ with $\E
\widetilde\xi_0^2 = \E \xi_0^2 = \frac1{2n+1}$, and set $ f_{\pm}
= \pm\widetilde\xi_0 b_x + f_x$. These are Gaussian spherical
harmonics having the same distribution as $f$. Note that
\[
f = \xi_0 b_x + \frac12 \left( f_+ + f_- \right)\,,
\]
and that by Claim~\ref{claim11}
\[
\PP \big( \{ \max_{\mathcal D(\rho, x)} |f_+| \le C_0 \} \cap \{
\max_{\mathcal D(\rho, x)} |f_-| \le C_0 \} \big) \ge 1 - (
\frac13 + \frac13) = \frac13\,.
\]

Now, consider the event $\Omega_x$ that $f(x)\ge C_0$ and
$f\big|_{\partial \mathcal D(x; \rho/n) }\le - C_0 $.  The event
$\Omega_x$ happens provided that
\[
\xi_0 \sqrt{n} \ge 2 c_1^{-1}C_0 \qquad {\rm and} \qquad
\max_{\mathcal D(\rho, x)} |f_\pm| \le C_0\,.
\]
Therefore,
\begin{multline*}
\PP (\Omega_x) \ge \PP ( \xi_0 \sqrt{n} \ge 2c_1^{-1}C_0)\, \cdot
\\ \cdot \PP \big( \{ \max_{\mathcal D(\rho, x)} |f_+| \le C_0\}
\cap \{ \max_{\mathcal D(\rho, x)} |f_-| \le C_0\} \big)  \ge
\kappa > 0\,.
\end{multline*}
Here, $\kappa$ is a positive numerical constant. (Recall that the
variance of the Gaussian random variable $\xi_0 \sqrt{n}$ is of
constant size.)

It remains to choose $\simeq n^2$ disjoint disks on $\Ss$ of
radius $2\rho/n$. Each of them contains a component of $Z(f)$ with
probability at least $\kappa$. Hence, $\E N(f) \gtrsim n^2$.
\hfill $\Box$

\section{Exponential concentration near the median}
\label{section_concentration}

\subsection{Main lemma}

We would like to use Levy's concentration of measure principle. To
this end, we need to show that the number $N(f)$ doesn't change
too much under slight perturbations of $f$. We won't be able to
prove it for all $f\in \HH_n$ but we will show that the
``unstable'' spherical harmonics $f$ for which small perturbations
can lead to a drastic decrease in the number of nodal lines are
exponentially rare. More precisely, we will prove the following

\begin{lemma}\label{lemma}
For every $\e>0$, there exists $\rho>0$ and an exceptional set
$E\subset \HH_n$ of probability $\PP(E)\le C(\e)e^{-c(\e)n}$ such
that for all $f\in \HH_n\setminus E$ and for all $g\in \HH_n$
satisfying $\|g\| \le \rho$, we have $N(f+g)\ge N(f)-\e n^2$.
\end{lemma}

Let us show that Lemma~\ref{lemma} ensures an exponential
concentration of $\dfrac{N(f)}{n^2}$ near its median $a_n$.
Consider first the set $F=\{f\in\HH_n \colon N(f)>(a_n+\e)n^2\}$.
Then for $ f\in (F\setminus E)_{+\rho}$, we have $N(f)>a_nn^2$,
and therefore, $\PP( (F\setminus E)_{+\rho} )\le \frac 12$. Hence,
$\PP(F\setminus E)\le 2e^{-c\rho^2 n}$ and
$$
\PP(F)\le 2 e^{-c\rho^2n}+C(\e)e^{-c(\e)n}\le C(\e)e^{-c(\e)n}\,.
$$
Now consider the set $G = \{f\in\HH_n\colon N(f)<(a_n-\e)n^2\}$.
Then
\[
G_{+\rho} \subset \{f\in \HH_n\,:\,N(f)<a_n^2\}\cup E
\]
and, thereby,
\[
\PP( G_{+\rho})\le \frac 12+ C(\e)e^{-c(\e)n}<\frac 34
\]
for large $n$ and it follows that $\PP( G ) \le 2e^{-c\rho^2 n}$
for large $n$. It remains to note that, for fixed $c(\e)$, we can
always make the estimate hold for small $n$ by increasing the
value of $C(\e)$. \hfill $\Box$

\subsection{Unstable spherical harmonics are exponentially rare}
The exceptional set $E$ of ``unstable spherical harmonics'' is
constructed as follows. We take a sufficiently large positive $R$
and cover the unit sphere $\Ss$ by approximately $R^{-2}n^2$
spherical disks $\mathcal D_j$ of radii $R/n$ with multiplicity of
covering bounded by a positive numerical constant. Let $3\mathcal
D_j$ be the disks of radii $3R/n$ with the same centers as
$\mathcal D_j$. Fix some small $\alpha,\beta>0$. We shall call a
disk $3\mathcal D_j$ \textit{stable} for a function $f\in \HH_n$
if there is no point $x\in 3\mathcal D_j$ such that $|f(x)|<
\alpha$ and $|\nabla f(x)|<\beta n$ simultaneously. Otherwise we
shall call the disk $3 \mathcal D_j$ unstable. Finally, fix a
small $\delta>0$. We shall call a function $f\in\HH_n$ {\em
exceptional} if the number of the unstable disks for this function
exceeds $\delta n^2$.

\bigskip
Our first task will be to find the conditions that would imply
that the exceptional functions are exponentially rare. To this
end, note that if we can find $\delta n^2$ unstable disks, we can
also find $c_2\delta n^2$ unstable disks that are $4/n$-separated.
Now, for each unstable disk $3\mathcal D_j$ in this well-separated
family, pick a point $x_j\in 3\mathcal D_j$ where $|f|<\alpha$ and
$|\nabla f|<\beta n$ simultaneously. Fix $\gamma\in(0,1)$ and
consider the disks $\mathcal D(x_j,\gamma/n)$. They are pairwise
disjoint. Let $\displaystyle M_j=\max_{\mathcal
D(x_j,\gamma/n)}|\nabla\nabla f|$. Note that
\[
\int_{\mathcal D(x_j, 2/n)} f^2 \gtrsim n^{-6}M_j^2
\]
and that the disks $\mathcal D(x_j,2/n)$ are also pairwise
disjoint. Hence,
$$
\sum_j M_j^2\le C_1n^6\|f\|^2.
$$
Now there are 2 possibilities: either $\|f\| > 2$, or for the
majority of our disks $3 \mathcal D_j$, we have $M_j\le C_2
\delta^{-1/2} n^2$.

The functions for which the first possibility holds are
exponentially rare (Claim~\ref{claim8}).

On the other hand, if the second possibility holds, we can
conclude using the Taylor formula that in at least $\tfrac12 c_2
\delta n^2$ pairwise disjoint disks of radius $\gamma/n$, we have
the estimates
$$
\begin{aligned}
|f|&\le \alpha+\beta\gamma + C_2\delta^{-1/2}\gamma^2
\\
|\nabla f|&\le \left( \beta + C_2\delta^{-1/2}\gamma \right) n
\end{aligned}
$$
Let now $g\in \HH_n$ satisfy $\|g\| \le \tau$. Then the number of
our disks where $\max|g|$ is much greater than $\delta^{-1/2}\tau$
or $\max|\nabla g|$ is much greater than $\delta^{-1/2}\tau n$ is
small compared to $\delta n^2$. Thus, we can conclude that $f+g\in
U$ where $U$ is the set of all $h\in\HH_n$ satisfying
$$
\mathcal A(h)=\operatorname{Area}\{x\in\Ss\,:\,|h(x)|\le A,
|\nabla h(x)|\le Bn \}\ge c_3\delta\gamma^2\,.
$$
with
$$
\begin{aligned}
A&=\alpha+\beta\gamma + C_3\delta^{-1/2}(\gamma^2+\tau)
\\
B&=\beta + C_3\delta^{-1/2}(\gamma+\tau)
\end{aligned}
$$
We want to show that $\PP(U)\le\frac 12$ and use Levy's
concentration of measure principle to conclude that the
probability that $f$ is exceptional does not exceed $2 e^{-c\tau^2
n}$. By independence of $h(x)$ and $\nabla h(x)$
(Claim~\ref{claim10}), we see that, for each $x\in \Ss$, we have
$$
\PP\{|h(x)|\le A, |\nabla h(x)|\le Bn\}\le C_4 AB^2\,.
$$
Due to rotation invariance, \[ \E\mathcal A(h)\le  \PP\{|h(x)|\le
A, |\nabla h(x)|\le Bn\} \le C_4 AB^2\,, \] and we can draw the
desired conclusion if $2 C_4 A B^2\le c_3\delta\gamma^2$. At this
point we shall just note that, for given $\delta>0$, we can always
choose some positive $\gamma$, $\alpha$, $\beta$ and $\tau$ to
satisfy this inequality just because the right hand side behaves
like $\gamma^2$ and the left hand side behaves like $\gamma^4$
when $\alpha=\beta=\tau=0$ and $\gamma\to 0+$. We shall postpone
the optimal choice of parameters until later when all the
relations between them will be discerned. \hfill $\Box$

\subsection{}
Now our task is to find the conditions that will ensure that
$N(f+g)\ge N(f)-\e n^2$ whenever $f$ is not exceptional and
$\|g\|\le \rho$. We need to estimate the number of components of
$Z(f)$ that may disappear or merge with some other components in
the process of perturbing $f$ by $g$.

First of all, we discard all components of $Z(f)$ whose diameters
is greater than $R/n$. Since the total length of $Z(f)$ does not
exceed $Cn$ (Claim~\ref{claim3}), we can conclude that the number
of such components is much less than $\e n^2$ if $R$ is much
greater than $\e^{-1}$.

Now, for each small component $\Gamma$, we fix the disk $\mathcal
D_j$ that intersects $\Gamma$. Then $\Gamma$ lies deeply within
the disk $3\mathcal D_j$: the distance from $\Gamma$ to the
boundary of $3\mathcal D_j$ is at least $R/n$.

Next, we forget about all small components whose disks are
unstable. The area estimate (Claim~\ref{claim4}) implies that each
unstable disk $3\mathcal D_j$ can contain at most $CR^2$ small
components, so, if $f$ is not exceptional, the total number of
small components whose disks are unstable does not exceed $\delta
R^2 n^2$, which is much smaller than $\e n^2$ if $\delta R^2$ is
much less than $\e$.

We need to show that if the disk corresponding to the component
$\Gamma$ is stable, then the component $\Gamma$ won't disappear or
merge with another component unless $\displaystyle \max_{3\mathcal
D_j} |g|\ge \alpha$. This will follow from the next claim which we
will use later in various contexts.

\begin{claim}\label{claim13a} Fix positive $\mu$
and $\nu$. Let $\mathfrak D$ be a disk and let $F$ be a
$C^1$-function on $\mathfrak D$ such that at each point
$x\in\mathfrak D$ either $|F(x)|>\mu$ or $|\nabla f(x)|>\nu$. Then
each component $\Gamma$ of the zero set $Z(F)$ with ${\rm dist}\,
(\Gamma, \partial\mathfrak D)>\mu/\nu$ is contained in an
``annulus'' $A_\Gamma$ bounded by two smooth curves and such that

\par\noindent (i) $F=+\mu$ on one boundary curve of $A_\Gamma$ and
$=-\mu$ on the other;

\par\noindent (ii) $A_\Gamma\subset \Gamma_{+\mu/\nu}$;

\par\noindent (iii) the annuli $A_\Gamma$ are pairwise disjoint.
\end{claim}

Note that if $G$ is an arbitrary continuous function on $\mathfrak
D$ with $\sup |G| < \mu$, then $A_\Gamma$ must contain at least
one component of the zero set $Z(F+G)$. We get

\begin{cor}\label{claim13} In the assumptions of the previous
claim, suppose that $G\in C(\mathfrak D)$ with $\sup |G| < \mu$.
Then each component $\Gamma$ of $Z(F)$ with ${\rm dist}\, (\Gamma,
\partial\mathfrak D)>\mu/\nu$ generates a component $\widetilde\Gamma$
of the zero set $Z(F+G)$ such that $\widetilde{\Gamma} \subset
\Gamma_{+\mu/\nu}$. Different components $\Gamma_1\ne\Gamma_2$ of
$Z(F)$ generate different components $\widetilde{\Gamma}_1\ne
\widetilde{\Gamma}_2$ of $Z(F+G)$.
\end{cor}

Later we'll use this corollary in various contexts.

\medskip\par\noindent{\em Proof of Claim~\ref{claim13a}:}
Replacing the function $F(u)$ by $\lambda_1 F(\lambda_2 u)$, we
may assume that $\mu=\nu=1$. This will simplify our notation.

Let us look at what happens with the connected component $\mathcal
F(t)$ of the set $\{|F|<t\}$ containing $\Gamma$ as $t$ increases
from $0$ to $1$. As long as $\mathcal F(t)$ stays away from the
boundary $\partial \mathfrak  D$, it cannot merge with another
component of $\{|F|<t\}$  because such a merge can occur only at a
critical point of $F$ and all critical values of $F$ in $\mathfrak
D$ are greater than $t$ in absolute value. For the same reason
neither of the two boundary curves of $\mathcal F(t)$ can collapse
and disappear. But $\mathcal F(t)$ cannot come too close to
$\partial \mathfrak D$ before it merges with some other component
either: indeed, if $x\in \mathcal F(t)$ and $\mathcal F(t)$ lies
at a positive distance from the boundary $\partial \mathfrak D$
then we can go from $x$ in the direction of $\nabla F$ if $F(x)<0$
and in the direction $-\nabla F$ if $F(x)>0$. In any case, since
$|\nabla F|>1$ in $\mathcal F(t)$, we shall reach the zero set
$Z(F)$ after going the unit length  or less. Since the only
component of $Z(F)$ in $\mathcal F(t)$ before any merges is
$\Gamma$, we conclude that $\mathcal F(t) \subset \Gamma_{+1}$.
Recalling that ${\rm dist}\, (\Gamma, \partial\mathfrak D)>1$, we
see that, for each $t\le 1$, $\mathcal F(t)$ stays away from the
boundary $\partial \mathfrak D$.

Thus, each component $\Gamma$ lies in an ``annulus'' $A_\Gamma =
\mathcal F(1)$ which is contained with its boundary in the open
disk $\mathfrak D$ and such that $F=1$ in one boundary curve of
$A_\Gamma$ and $F=-1$ on the other. By construction, the annuli
$A_\Gamma$  are pairwise disjoint. This proves the claim. \hfill
$\Box$

\medskip Now, we apply Corollary~\ref{claim13} to the functions
$F=f$ and $G=g$ on the disk $\mathfrak D = 3\mathcal D_j$ with
$\mu=\alpha$, and $\nu =\beta n$. We require that $\alpha/\beta <
R$. This guarantees that if $\Gamma$ is a component of $Z(f)$ with
${\rm diam}\, (\Gamma)\le R/n$ and $\Gamma\cap \mathcal D_j \ne
\emptyset$, then ${\rm dist}\, \big(\Gamma, \partial (3\mathcal
D_j) \big) \ge R/n > \alpha/(\beta n)$. We see that the only small
components of $Z(f)$ in stable disks $\mathcal D_j$ that can be
destroyed by perturbation of $f$ by $g$ are those that correspond
to the disks where $\displaystyle \max_{3\mathcal D_j}|g|\ge
\alpha$. By the mean value property (Claim~\ref{claim2}), the
number of such disks does not exceed $C\rho^2\alpha^{-2}n^2$ and,
by the area estimate (Claim~\ref{claim4}), the number of the
corresponding components is bounded by $C\rho^2\alpha^{-2}R^2
n^2$, which is much less than $\e n^2$ if $\rho^2$ is much less
than $\e\alpha^2R^{-2}$.

\subsection{Tuning the parameters}
Now it is time to make the choice of our parameters. First, let us
list the constraints introduced above:
\[
\rho^2 \ll \epsilon \alpha^2 R^{-2}, \qquad \alpha \ll R\beta,
\qquad \delta R^2 \ll \epsilon, \qquad R\gg \epsilon^{-1}\,,
\]
and
\[
\big( \alpha+\beta\gamma + \delta^{-1/2}(\gamma^2+\tau) \big)
\big( \beta + \delta^{-1/2}(\gamma+\tau) \big)^2 \ll \delta
\gamma^2\,.
\]

We take
\[
R \simeq \e^{-1}, \quad \delta \simeq \e^3, \quad  \rho^2 \simeq
\alpha^2 \epsilon^3, \quad {\rm and} \quad  \beta \simeq \e\alpha.
\]
The quantity we want to maximize is
$c(\e)\simeq\min(\rho^2,\tau^2) \simeq \min(\tau^2,\alpha^2\e^3)$
subject to the constraint
$$
[\alpha+\e^{-3/2}(\gamma^2+\tau)] \cdot
[\e^2\alpha^2+\e^{-3}(\gamma^2+\tau^2)] \ll \e^3\gamma^2
$$
(we neglected absolute constants and the term
$\beta\gamma\simeq\alpha\e\gamma<\alpha$ in the first bracket).
Denoting the minimum to maximize by $m$, we see that we have to
put $\tau=m^{1/2}$, $\alpha=m^{1/2}\e^{-3/2}$. This leads to the
constraint
$$
\e^{-3/2}[m^{1/2}+\gamma^2]\cdot\e^{-3}[\gamma^2+m] \ll
\e^3\gamma^2\,.
$$
Again, we neglected $\e^2\alpha^2=\e^{-1}m<\e^{-3}m$. Rewrite this
constraint as
$$
[m^{1/2}+\gamma^2]\cdot[m+\gamma^2] \ll \e^{15/2}\gamma^2\,.
$$
It is immediate from here that $m\ll \e^{15}$. On the other hand,
taking $\gamma^2\simeq\e^{15/2}$, we see that this upper bound can
be attained. Thus, the proof we presented gives $c(\e) \simeq
\e^{15}$. \hfill $\Box$

\section{Existence of the limit $\displaystyle \lim_{n\to\infty}a_n$}
\label{section_limit}

In this section, we denote the spherical harmonics from $\HH_n$ by
$f_n$. Since the random variable $N(f_n)/n^2$ exponentially
concentrates near its median $a_n$ and is uniformly bounded, it
suffices to show that the sequence of means $\big\{ \E N(f_n)/n^2
\big\}$ converges. Then the sequence of medians $\big\{a_n\big\}$
converges to the same limit. In what follows, we'll show that
$\big\{ \E N(f_n)/n^2 \big\}$ is a Cauchy's sequence.

\subsection{Some integral geometry}
Let $\mathcal G$ be a system of $N(\mathcal G)$ loops on the
sphere $\Ss$. By $N_*(\mathcal G, \mathcal D)$ we denote the
number of loops from $\mathcal G$ that are contained in the
spherical disk $\mathcal D$, and by $N(\mathcal G, \mathcal D)$ we
denote the number of loops from $\mathcal G$ that intersect
$\mathcal D$. We fix $\rho>0$ and denote $\mathcal D_x = \mathcal
D(x, \rho)$, $S = \sigma(\mathcal D_x)$. Note that the area $S$
does not depend on $x$.

\begin{claim}\label{claim_IG}
\[
\frac1{S} \int_{\Ss} N_*(\mathcal G, \mathcal D_x)\, d\sigma (x)
\le N(\mathcal G) \le \frac1{S} \int_{\Ss} N(\mathcal G, \mathcal
D_x)\, d\sigma (x)\,.
\]
\end{claim}

\medskip\par\noindent{\em Proof:} Fix a loop $\Gamma\in \mathcal
G$ and note that
\[
\sigma\big( \{x\colon \Gamma\subset\mathcal D_x\}\big) \le S\,
\quad {\rm and} \quad S \le \sigma\big( \{x\colon \Gamma \cap
\mathcal D_x \ne\emptyset \} \big)\,.
\]
To prove the first inequality, we fix an arbitrary point
$y\in\Gamma$ and observe that  $\{x\colon \Gamma\subset\mathcal
D_x \} \subset \mathcal D_y$. Similarly, to prove the second
inequality holds, we fix a point $y\in\Gamma$ and note that
$\{x\colon \Gamma\cap \mathcal D_x \ne\emptyset\} \supset \mathcal
D_y$. \hfill $\Box$

\medskip Now, we fix $1 \ll d \ll R$, put $\rho = R/n$, and let $n$
go to $\infty$. We set
\[
N(f_n, \mathcal D_x) \stackrel{\rm def}= N( Z(f_n), \mathcal D_x)
\quad {\rm and} \quad  N_*(f_n, \mathcal D_x) \stackrel{\rm def}=
N_*( Z(f_n), \mathcal D_x)\,.
\]
We call the component $\Gamma$ of $Z(f_n)$ $d$-{\em normal} if its
diameter does not exceed $d$, and denote by $N_d(f_n, \mathcal
D_x)$ the number of $d$-normal components of $Z(f_n)$ that
intersect the disk $\mathcal D_x$.

By Claim~\ref{claim3}, the total number of $d$-abnormal components
does not exceed $Cn^2/d$. Thus, applying Claim~\ref{claim_IG} to a
spherical harmonic $f_n$, we get
\begin{multline*}
\frac1{Sn^2} \int_{\Ss} N_*(f_n, \mathcal D_x)\,
d\sigma (x) \\
\le \frac{N(f_n)}{n^2} \le \frac{N_d(f_n)}{n^2} + \frac{C}{d} \\
\le \frac1{Sn^2} \int_{\Ss} N_d(f_n, \mathcal D_x)\, d\sigma (x) +
\frac{C}{d}\,.
\end{multline*}
Taking the expectation and using rotation invariance of the
distribution of random spherical harmonics (and recalling that
$\sigma(\Ss)=1$), we continue our chain of estimates
\begin{equation}\label{eq_IG}
\frac{\E N_* (f_n, \mathcal D_{x_0})}{Sn^2} \le \frac{\E
N(f_n)}{n^2} \le \frac{\E N_d (f_n, \mathcal D_{x_0})}{Sn^2} +
\frac{C}{d}\,,
\end{equation}
where $x_0$ is an arbitrary point on $\Ss$.

\subsection{Scaling}
We fix a point $x_0\in\Ss$, denote by $x_0^*$ the antipodal point,
and by $\pi_{x_0}\colon \Ss\setminus \{x_0^*\} \to T_{x_0}\Ss$ the
stereographic projection ($\pi_{x_0} (x_0) = 0$), and define a
function $F_n$ on $T_{x_0}\Ss$ by $\displaystyle F_n(u) = \big(
f_n \circ \pi_{x_0}^{-1})(\frac{u}{n} \big)$. We also set
$D(t)=D(0, t)$. By $N_d(F_n, R)$ we denote the number of
components of the nodal set $\{F_n=0\}$ of diameter at most $d$
that intersect the disk $D(R)$, and by $N_*(F_n, R)$ we denote the
number of components that are contained in the disk $D(R)$. Note
that $\pi_{x_0} (\mathcal D_{x_0}) = D(0, \widetilde{R}/n)$ with
some $\widetilde{R}>R$. Then $ N_*(F_n, \widetilde{R}) = N_*(f_n,
\mathcal D_{x_0}) $. Since $1\ll d\ll R \ll n$, we have $N_d(f_n,
\mathcal D_{x_0}) \le N_{2d}(F_n, \widetilde{R})$,
\[
1< \frac{\widetilde{R}}{R} < 1 + C \Big( \frac{R}{n} \Big)^2,
\quad {\rm and} \quad 1- C \Big( \frac{R}{n} \Big)^2 <
\frac{Sn^2}{R^2} < 1\,.
\]
Then, using that $\E N_d(f_n, \mathcal D_{x_0}) \lesssim S$, we
get a scaled version of \eqref{eq_IG}:
\[
\frac{\E N_*(F_n, \widetilde{R})}{ \widetilde{R}^2}  \le \frac{\E
N(f_n)}{n^2} \le \frac{\E N_{2d}(F_n,
\widetilde{R})}{\widetilde{R}^2} + \frac{C\widetilde{R}^2}{n^2} +
\frac{C}{d}\,,
\]
valid for $1\ll d\ll \widetilde{R} \ll n$. At this point we
simplify our notation returning to notation $d$ instead of $2d$
and to $R$ instead of $\widetilde R$. We get
\begin{claim}\label{claim_prep} For any $d$ and $R$ such that
$1\ll d\ll R \ll n$, we have
\[
\frac{\E N(f_n)}{n^2} - \frac{\E N(f_m) }{m^2}  \le \frac{\E
\left\{ N_d (F_n, R) - N_*(F_m, R) \right\}}{R^2} +
\frac{CR^2}{n^2} + \frac{C}{d}\,.
\]
\end{claim}

\medskip
Later, estimating the expectation on the right-hand side, we'll
use that the expression $\displaystyle \frac{N_d (F_n, R) -
N_*(F_m, R) }{R^2}$ is bounded from above by a positive numerical
constant. This follows from

\begin{claim}\label{claim_UB} We have
\[
\frac{N_d (F_n, R)}{R^2} \le C\,,
\]
uniformly with respect to $R$ and $n$.
\end{claim}

\medskip\par\noindent{\em Proof:} Obviously, $N_d (F_n, R) \le N_* (F_n,
R+d)$. By scaling Claim~\ref{claim4}, the area of each nodal
domain of the function $F_n$ cannot be less than a positive
numerical constant $c_4$, therefore, $N_* (F_n, R+d) \le \pi
(R+d)^2 c_4^{-1}$. Hence, the claim. \hfill $\Box$

\bigskip In what follows, we show that if we discard some events of small
probability, the difference $N_d(F_n, R) - N_* (F_m, R)$ will be
small. In view of Claims~\ref{claim_prep} and \ref{claim_UB}, this
will prove that $\E N(f_n)/n^2$ is a Cauchy's sequence.

The main idea is to show first that if $m$ and $n$ are
sufficiently large, then the function $F_m$ can be viewed as a
statistically small $C^1$-perturbation of the function $F_n$, and
therefore, outside of small events, $N_d(F_n, R)$ cannot be much
larger than $N_*(F_m, R)$. We start with
\begin{claim}\label{claim_conv}
Given a finite set of points $\big\{ u_j\big\}$, the random
vectors $F_n(u_j)$ converge in distribution as $n\to\infty$.
\end{claim}

\medskip\par\noindent{\em Proof:} We use Claim~\ref{claim5.5}:
\[
\E \big\{ f_n (x) f_n (y) \big\} = \frac1{2n+1} \sum_{k=-n}^n
Y_k(x) Y_k(y) = P_n \big(\cos \Theta (x, y) \big)
\]
where $\Theta (x, y)$ is the angle between $x$ and $y$ as vectors
in $\mathbb R^3$. Then  the scaled covariance equals
\begin{multline*}
\E \big\{ F_n (u) F_n (v) \big\} = \E \Big\{ (f_n \circ
\pi_{x_0}^{-1}) \big( \frac{u}{n} \big) (f_n\circ \pi_{x_0}^{-1})
\big( \frac{v}{n} \big) \Big\} \\
= P_n \big( \cos\Theta ( \pi_{x_0}^{-1}\big(\frac{u}{n}\big),
\pi_{x_0}^{-1} \big( \frac{v}{n})\big)\, \big)\,.
\end{multline*}
When $n$ goes to $\infty$, the angle between the points
$\pi_{x_0}^{-1}\big(\tfrac{u}{n}\big)$, and  $\pi_{x_0}^{-1} \big(
\tfrac{v}{n}\big)$ on the sphere is equivalent to $|u-v|/n$
(locally uniformly in $u$ and $v$). Therefore, by Hilb's theorem
(Claim~\ref{claim7}), the scaled covariance $ \E \big\{ F_n (u)
F_n (v) \big\} $ converges to the Bessel kernel $J_0 (|u-v|)$
locally uniformly in $u$ and $v$.

Recall that the vector $\big\{ F_n(u_j) \big\}$ is a Gaussian one,
and that the convergence of covariance matrices of a sequence of
Gaussian vectors yields convergence in distribution of the
vectors. \hfill $\Box$

\subsection{Discarding small events}

\subsubsection{}

Consider the event
\[
\Omega_n^{(1)} = \Big\{ \int_{D(5R)} F_n^2\, dm > R^3 \Big\}\,.
\]
Since at any point $x\in\Ss$, $\E |f_n(x)|^2 = 1$, we have
\[
\E \int_{D(5R)} F_n^2\, dm  = \int_{D(5R)} \E F_n^2 \, dm =
CR^2\,.
\]
Then, by Chebyshev's inequality, $\PP (\Omega_n^{(1)}) \lesssim
R^{-1}$ and $\PP \big(\Omega_n^{(1)} \cup \Omega_m^{(1)} \big)
\lesssim R^{-1}$. Throwing away these events, we assume that
\[
\max\Big\{ \int_{D(5R)} F_n^2, \int_{D(5R)} F_m^2 \Big\} \le
R^3\,.
\]
By Claim~\ref{claim2} this yields the estimates
\begin{equation}\label{eqE}
\| F_n \|_{C^2(D(4R))}\,, \ \| F_m \|_{C^2(D(4R))} \lesssim
R^{3/2}\,.
\end{equation}

\subsubsection{}

Now, we fix a finite $R^{-(a+2)}$-net $\{u_j\}$ in the disk
$D(4R)$. The parameter $a>1$ will be chosen later. Since by
Claim~\ref{claim_conv},
\[
\lim_{\min(m, n) \to\infty} \PP \big( \max_j | (F_n - F_m)(u_j)| >
\epsilon \big) = 0\,,
\]
in what follows, we discard the event
\[
\Omega_{n, m}^{(2)} = \Big\{ \max_{j} \big| (F_n - F_m)(u_j)
\big|
> \frac1{R^{a+2}} \Big\}
\]
and assume that
\[
\max_{j} \big| (F_n - F_m)(u_j)  \big| \le \frac1{R^{a+2}}\,.
\]
Using a priori estimates \eqref{eqE}, we get
\[
\max_{D(4R)} \big| F_n - F_m \big| \le \frac1{R^{a+2}} +
\frac{CR^{3/2}}{R^{a+2}} < \frac{C}{R^{a+1/2}}\,.
\]
Then, scaling local gradient estimates from Claim~\ref{claim2}, we
get
\[
\max_{D(3R)} \big| \nabla (F_n - F_m) \big| \le
\frac{C}{R^{a+1/2}}  \ll \frac1{R^a}
\]
if $R$ is big enough. We conclude that
\[
\| F_n - F_m \|_{C^1(D(3R))} \le  \frac1{R^a}\,,
\]
that is, outside of events $\Omega_n^{(1)} \cup \Omega_m^{(1)}$,
and $\Omega_{n, m}^{(2)}$, the random function $F_m$ indeed can be
viewed as a small $C^1$-perturbation of the random function $F_n$
in the disk $D(3R)$.

\subsubsection{}

To be sure that a $R^{-a}$-perturbation of the function $F_n$ does
not decrease drastically the number of the components of the zero
set $\big\{ F_n=0 \big\}$ in the disk $D(R)$, we need to know that
the function $F_n$ is ``stable'' in a larger disk $D(3R)$, e.g.,
that
\[
\min_{D(3R)} \big\{ |F_n| + |\nabla F_n| \big\} > \frac2{R^a}\,.
\]

Suppose that
\[
\min_{D(3R)} \big\{ |F_n| + |\nabla F_n| \big\} \le \frac2{R^a}\,,
\]
and estimate the probability of this event (we call it
$\Omega_n^{(3)}$). We fix a $R^{-(a+2)}$-net $\{u_j\}$, this time
in the disk $D(3R)$, that contains at most $CR^{2a+6}$ elements.

Suppose that, at some point $u\in D(3R)$,
\[
|F_n(u)| + |\nabla F_n(u)| \le \frac2{R^a}\,.
\]
Then there is a point $u_\ell$ of our net  such that
\begin{equation}\label{eq*}
|F_n(u_{\ell})| + |\nabla F_n (u_{\ell}) | \le \frac2{R^a} +
\frac{R^{3/2}}{R^{a+2}} < \frac3{R^a}
\end{equation}
(we again used a priori estimates \eqref{eqE}).

By the independence Claim~\ref{claim10}, the probability that  in
a {\em given point} $u_j$ from our net condition \eqref{eq*} holds
does not exceed $CR^{-3a}$. Hence, the probability that
\eqref{eq*} holds at {\em some point} of the net does not exceed
$CR^{2a+6} \cdot CR^{-3a} = CR^{6-a}$ and tends to $0$ as
$R\to\infty$ provided that $a>6$. Hence, choosing $a=7$, we
achieve that $\PP ( \Omega_n^{(3)}) \le CR^{-1}$.

\bigskip To summarize, we denote by $\Omega^*$ the {\em complement} to
the union of our small events $ \Omega_n^{(1)} \cup \Omega_m^{(1)}
\cup \Omega_{n, m}^{(2)} \cup \Omega_n^{(3)}$. Then
\[
\PP (\Omega^* \ {\rm doesn't\ occur}\,) \le \frac{C}{R} +
\kappa(n, m)\,, \qquad \lim_{\min(n, m)\to\infty} \kappa(n, m) =
0\,.
\]
We have proved the following
\begin{claim}\label{claim.summary} Given $R$, $n$ and $m$ such that
$1\ll R\ll\min(n, m)$, there exists an  event $\Omega^*$ such that
if it happens then
\[
\| F_n - F_m\|_{C^1(D(3R))} \le \frac1{R^7}\,,
\]
\[
\min_{D(3R)} \big\{ |F_n| + |\nabla F_n| \big\} > \frac2{R^7}\,,
\]
and
\begin{multline*}
\frac{\E N(f_n)}{n^2} - \frac{\E N(f_m) }{m^2}
\\ \lesssim \sup_{\Omega^*\ {\rm occurs}}\, \frac{N_d (F_n, R)
- N_*(F_m, R)}{R^2} + \frac{R^2}{n^2} + \frac1{d} + \kappa(n,
m)\,,
\end{multline*}
with
\[
\lim_{\min(n, m)\to\infty} \kappa(n, m) = 0, \quad {\rm and} \quad
1\ll d \ll R.
\]
\end{claim}

\subsection{}

The following claim estimates the supremum on the right-hand side
of the previous bound.

\begin{claim}\label{claim_final}
If the event $\Omega^*$ occurs, then
\[
\frac{N_d(F_n, R) - N_* (F_m, R)}{R^2} \le \frac{Cd}{R}\,.
\]
\end{claim}

\medskip\par\noindent{\em Proof:} First, note that
$N_d(F_n, R) - N_* (F_m, R)$ does not exceed the number of
components of the zero set $\big\{ F_m=0\big\}$ that are contained
in the annulus $A = \big\{ R-d-2 \le |u| \le R+d+2 \big\}$. To see
this, we apply Corollary~\ref{claim13} to the functions $F=F_n$
and $G=F_m-F_n$ with $\mathfrak D = D(3R)$, and $\mu = \nu =
R^{-7}$. By this Corollary, each $d$-normal component $\Gamma$ of
$Z(F_n)$ such that $\Gamma \cap D(R) \ne \emptyset$ generates a
component $\widetilde \Gamma$ of the zero set $Z(F_m)$ such that
$\widetilde{\Gamma} \subset \Gamma_{+1}$. If $\widetilde\Gamma$ is
not contained in $D(R)$ then, observing that the diameter of
$\widetilde\Gamma$ does not exceed $d+2$, we conclude that it must
be contained in the annulus $A$.

Since the area of each nodal domain of $F_m$ cannot be less than a
positive numerical constant (Claim~\ref{claim4}), we see that the
number of components of the zero set $\big\{ F_m=0\big\}$ that are
contained in the annulus $A$ cannot exceed
\[
C \big[ (R+d+2)^2-(R-d-2)^2) \big] \le CdR\,,
\]
proving the claim. \hfill $\Box$

\bigskip Combining Claims~\ref{claim.summary} and
\ref{claim_final}, we obtain
\[
\frac{\E N(f_n)}{n^2} - \frac{\E N(f_m) }{m^2} \lesssim
\frac{d}{R} + \frac1{d} + \frac{R^2}{n^2}  + \kappa(n, m)\,,
\]
First, we set $d=\sqrt{R}$. Then, given $\epsilon>0$, we choose
$R$ so big that $1/\sqrt{R}< \epsilon$. At last, we choose $n$ and
$m$ so large that $\kappa (n, m) < \epsilon$ and
$R^2/n^2<\epsilon$. Then we get
\[
\frac{\E N(f_n)}{n^2} - \frac{\E N(f_m) }{m^2} \lesssim
\epsilon\,.
\]
This completes the proof of convergence of $\E N(f_n)/n^2$, and
hence finishes off the proof of the theorem. \hfill $\Box$

\section{Sharpness of Theorem~\ref{thm.main}: $\PP \left\{ N(f)
< \kappa n^2\right\} \ge e^{-C(\kappa) n}$}\label{section_example}

The idea is very simple: the zero set of the zonal spherical
harmonic $Y_0$ is a union of $n$ circles of constant latitude. On
the other hand, by Claim~\ref{claim_zonal}, the zonal harmonic
$Y_0$ is stable, and  therefore, its small $L^2$-perturbations
cannot increase much the number of components of the nodal set.

\medskip
Let
\[
f = \sum_{k=-n}^n \xi_k Y_k\,, \qquad \E \xi_k^2 = \frac1{2n+1}
\]
be a Gaussian spherical harmonic of degree $n$. Consider the event
\[
\Omega = \big\{ \xi_0^2 \ge 1, \quad \sum_{k\ne 0} \xi_k^2 \le
\rho^2 \big\}\,,
\]
where $\rho$ is a small positive constant which we shall choose
later. We have
\[
\PP \big\{ \Omega \big\} = \PP \big\{ \xi_0^2 \ge 1 \big\} \cdot
\PP \big\{ \sum_{k\ne 0} \xi_k^2 \le \rho^2 \big\} \ge
e^{-C(\rho)n}\,.
\]
In what follows, we assume that the event $\Omega $ occurs. Then
$f=\xi_0 Y_0 + g$ with $\|g\| \le \rho$.

Again, we cover the sphere $\Ss$  by $\simeq R^{-2} n^2 $
spherical disks $\mathcal D_j$ of radius $R/n$ with $R$ (depending
on $\kappa$) to be chosen later. The disk $\mathcal D_j$ is {\em
good} if $\displaystyle \max_{3\mathcal D_j} \big( |g| +
\tfrac1{n} |\nabla g| \big) < \tfrac14 c_0$ where $c_0$ such a
constant that $\displaystyle (\xi_0 Y_0)^2 + \frac1{n^2} |\nabla
(\xi_0 Y_0)|^2 \ge c_0^2$ everywhere on $\Ss$
(Claim~\ref{claim_zonal}). Comparing the areas
(Claim~\ref{claim2}), we see that the number of {\em bad} disks is
$\lesssim c_0^{-2} \rho^2  n^2 $.

Now, let $\Gamma$ be a connected component of the nodal set
$Z(f)$. Then at least one of the three following possibilities
must occur:

\smallskip\par\noindent({\bf i}) The component $\Gamma$ has diameter larger than $R/n$.

\smallskip\par\noindent({\bf ii}) The component $\Gamma$ has diameter less than $R/n$ and
intersects a good disk $\mathcal D_j$.

\smallskip\par\noindent({\bf iii}) The component $\Gamma$ has diameter less than $R/n$ and
intersects a bad disk $\mathcal D_j$.

\smallskip

By Claim~\ref{claim3}, the number of components of the first type
is bounded by $CR^{-1} n^2 \ll \kappa n^2$, provided that $R$ is
chosen sufficiently large.

If the disk $\mathcal D_j$ is good, then, for $x\in 3\mathcal
D_j$, we cannot have $|f(x)|< \tfrac14 c_0$ and $|\nabla f(x)|<
\tfrac{n}4 c_0$ at the same time. Therefore, we can apply
Corollary~\ref{claim13} to the functions $F=f$ and $G=-g$ with
$\mu =\tfrac14 c_0$ and $\nu = \tfrac{n}4 c_0$. We see that each
component $\Gamma$ of the zero set $Z(f)$ of the second type
generates a component $\widetilde\Gamma$  of the zero set $Z(Y_0)$
of diameter at most $(R+2)/n$. Recall that $Y_0$ is a zonal
spherical harmonic and its nodal set consists of the spherical
circumferences of constant latitude that are generated by zeroes
of Legendre polynomials. By Claim~\ref{claim6}, the components
$\widetilde\Gamma$ of diameter at most $(R+2)/n$ must be located
in a neighbourhood of one of the Poles, and there is only a
bounded number of them. Hence, the number of components $\Gamma$
of the second type remains bounded as $n$ goes to $\infty$.

At last, all components of the third type are contained in the set
$\displaystyle \bigcup_{\mathcal D_j {\rm\ is\ bad\ }} 3\mathcal
D_j$ of area $ \lesssim \rho^2 R^2 $, and by the area estimate
(Claim~\ref{claim4}) the number of such components is $\lesssim
\rho^2 R^2 n^2 \ll \kappa n^2$ provided that $\rho$ is properly
chosen. \hfill $\Box$

\end{document}